\documentclass[12pt]{article} % default is 10pt
\usepackage[utf8x]{inputenc} % to set input encoding (not needed with XeLaTeX)
\usepackage[T1]{fontenc} % to set font encoding in latex; difference noticed for <>
\usepackage{amsmath,amssymb,bbm,bm,cancel,commath,mathtools,wasysym,xfrac} % for mathematical symbols
\usepackage{graphicx} % for \includegraphics command and options
\usepackage{color,xparse} % to use color and to link arXiv properly
\usepackage{cite} % for compact citations
\usepackage{geometry} % to change the page dimensions
\usepackage{fancyhdr} % This should be set AFTER setting up the page geometry
\usepackage{setspace} % to set spacing in the document
\usepackage[nottoc,notlof,notlot]{tocbibind} % to put the bibliography in the ToC
\usepackage[unicode]{hyperref} % for bookmarks in Acrobat Reader and using hyperlinks
\hypersetup{bookmarksnumbered=true, bookmarksopen=true, bookmarksopenlevel=2, breaklinks=true, citecolor=blue, colorlinks=true, linkcolor=red, linktoc=page, pdfborder={0 0 0}, pdfstartview=FitH, plainpages=false, unicode=true, urlcolor=blue}

\DeclareUnicodeCharacter{"0393}{\Gamma}
\DeclareUnicodeCharacter{"0394}{\Delta}
\DeclareUnicodeCharacter{"0398}{\Theta}
\DeclareUnicodeCharacter{"039B}{\Lambda}
\DeclareUnicodeCharacter{"039E}{\Xi}
\DeclareUnicodeCharacter{"03A0}{\Pi}
\DeclareUnicodeCharacter{"03A3}{\Sigma}
\DeclareUnicodeCharacter{"03A5}{\Upsilon}
\DeclareUnicodeCharacter{"03A6}{\Phi}
\DeclareUnicodeCharacter{"03A8}{\Psi}
\DeclareUnicodeCharacter{"03A9}{\Omega}
\DeclareUnicodeCharacter{"03B1}{\alpha}
\DeclareUnicodeCharacter{"03B2}{\beta}
\DeclareUnicodeCharacter{"03B3}{\gamma}
\DeclareUnicodeCharacter{"03B4}{\delta}
\DeclareUnicodeCharacter{"03B5}{\epsilon}
\DeclareUnicodeCharacter{"03B6}{\zeta}
\DeclareUnicodeCharacter{"03B7}{\eta}
\DeclareUnicodeCharacter{"03B8}{\theta}
\DeclareUnicodeCharacter{"03D1}{\vartheta}
\DeclareUnicodeCharacter{"03B9}{\iota}
\DeclareUnicodeCharacter{"03BA}{\kappa}
\DeclareUnicodeCharacter{"03BB}{\lambda}
\DeclareUnicodeCharacter{"03BC}{\mu}
\DeclareUnicodeCharacter{"03BD}{\nu}
\DeclareUnicodeCharacter{"03BE}{\xi}
\DeclareUnicodeCharacter{"03C0}{\pi}
\DeclareUnicodeCharacter{"03C1}{\rho}
\DeclareUnicodeCharacter{"03C3}{\sigma} %σ
\DeclareUnicodeCharacter{"03C4}{\tau}
\DeclareUnicodeCharacter{"03C5}{\upsilon}
\DeclareUnicodeCharacter{"03C6}{\phi}
\DeclareUnicodeCharacter{"03D5}{\varphi}
\DeclareUnicodeCharacter{"03C7}{\chi}
\DeclareUnicodeCharacter{"03C8}{\psi}
\DeclareUnicodeCharacter{"03C9}{\omega}
\DeclareUnicodeCharacter{"21D0}{\Leftarrow}
\DeclareUnicodeCharacter{"0212B}{\AA}
\DeclareUnicodeCharacter{"00B7}{\cdot}
\DeclareUnicodeCharacter{"00B0}{^{\circ}}
\DeclareUnicodeCharacter{"266A}{\eighthnote}
\DeclareUnicodeCharacter{"266B}{\twonotes}
\PrerenderUnicode{ä}\PrerenderUnicode{×}
\PrerenderUnicode{₄}\PrerenderUnicode{₅}
\PrerenderUnicode{₆}\PrerenderUnicode{₇}
\PrerenderUnicode{→}\PrerenderUnicode{̃}

\newcommand{\Title}[1] {\title{\Huge #1}}
\newcommand{\TPheader}[3] {\date{}\maketitle\thispagestyle{fancy}\pagenumbering{alph}\lhead{#1}\chead{#2}\rhead{#3}\cfoot{}}
\newcommand{\makepage}[1] {\newpage\pagenumbering{#1}}

\newcommand{\Abstract}[1] {\begin{abstract}\normalsize #1 \end{abstract}}

\renewcommand{\appendix}{\setcounter{section}{0}\renewcommand{\thesection}{\Alph{section}}\renewcommand*{\theHsection}{app.\the\value{section}}} % Hsection command for correct hyperlinking in .pdf!
\newcommand\references[1]{\bibliographystyle{hephys}\bibliography{#1}}

\newcommand\eqs[1] {\begin{align}#1\end{align}}

\newcommand\eqst[1] {\begin{multline}#1\end{multline}}

\newcommand\equ[1] {\begin{equation}#1\end{equation}}

\newcommand\s {\sigma}

\renewcommand\( {\left(}
\renewcommand\) {\right)}

\newcommand\wt {\widetilde}

\DeclareMathOperator{\Vol}{Vol}

\newcommand\I {{\mathcal I}}

\newcommand\N {{\mathcal N}} 
\renewcommand\O {{\mathcal O}}

\newcommand\bZ {{\mathbb Z}}
\newcommand\fg {{\mathfrak g}}

\newcommand\nn {\nonumber\\}

\numberwithin{equation}{section} % Equations numbered as <section>.#
\begin{document}
\Title{Universal Pieces of Holographic Entanglement Entropy and Holographic Subregion Complexity}

\author{Sunandan Gangopadhyay$^{γ}$\footnote{\href{mailto:sunandan.gangopadhyay@bose.res.in}{sunandan.gangopadhyay@bose.res.in}}$~$, Dharmesh Jain$^{ψ}$\footnote{\href{mailto:dharmesh.jain@bose.res.in}{dharmesh.jain@bose.res.in}}$~$ and Ashis Saha$^{\s}$\footnote{\href{mailto:ashisphys18@klyuniv.ac.in}{ashisphys18@klyuniv.ac.in}}\bigskip\\
\emph{\normalsize ${}^{γ,ψ}$Department of Theoretical Sciences, S. N. Bose National Centre for Basic Sciences}\\
\emph{\normalsize Block--JD, Sector--III, Salt Lake City, Kolkata 700106, India} \medskip\\
\emph{\normalsize ${}^{\s}$Department of Physics, University of Kalyani, Kalyani 741235, India}
}

\TPheader{}{\today}{} %\TPheader{Date}{...}{PREPRINT}

\Abstract{We propose that the definition of holographic subregion complexity (HSC) needs a slight modification for supergravity solutions with warped anti-de Sitter (AdS) factors. Such warp factors can arise due to the nontrivial dilaton profile, for example, in $AdS_6$ solutions of type IIA supergravity. This modified definition ensures that the universal piece of the HSC is proportional to that of the holographic entanglement entropy, as is the case for supergravity solutions without warp factors. This also means that the leading behaviour at large $N$ is the same for both these quantities, as we show for some well-known supergravity solutions (with and without warp factors) in various dimensions. We also show that this relation between the universal pieces suggests ``universal'' relations between field theoretical analogue of HSC and the sphere partition function or Weyl $a$-anomaly in odd or even dimensions, respectively.
}

\makepage{arabic} %Comment this line to make ToC appear on the title page.
\tableofcontents
%\makepage{arabic}
%%%%%%%%%%%%%%%%%%%%%%%%%%%%%%%%%%%%%%%%%%%%%

\section{Introduction}
The gauge/gravity correspondence inspired holographic observations have been a matter of great interest as they provide a means to study and uncover fascinating features of strongly coupled field theories via their gravity duals\cite{Maldacena:1997re,Witten:1998qj,Aharony:1999ti}. In the landscape of quantum information theory, the holographic computations of entanglement entropy (EE) and quantum complexity (QC) of a conformal field theory (CFT) are prime applications of this correspondence\cite{Ryu:2006bv,Ryu:2006ef,Susskind:2014rva,Stanford:2014jda,Brown:2015bva,Brown:2015lvg,Alishahiha:2015rta}.

\paragraph{Entanglement Entropy.} The EE is a ``good measure'' of quantum entanglement for a pure quantum state\footnote{In case of mixed states, (logarithmic) negativity is a better candidate to quantify quantum entanglement; see, for example, \cite{Vidal:2002zz}.} and represents the amount of information stored in a quantum system. When a system can be divided into two subsystems, $A$ and $B$, then the definition for the EE of subsystem $A$ ($S_A$) follows along the lines of von Neumann entropy:
\equ{S_{A} = -tr[\rho_A \log\rho_A]\,,
}
where $\rho_A=tr_B [\rho_{total}]$ is the reduced density matrix of the subsystem $A$ obtained by tracing out the degrees of freedom of the subsystem $B$ from the total density matrix $\rho_{total}$ of the whole system. The holographic computation of EE of a CFT in $d$-dimensional spacetime is provided by the Ryu-Takayanagi (RT) prescription which is defined as \cite{Ryu:2006bv,Ryu:2006ef}
\equ{S_A=\frac{\mathrm{Area}(\gamma_A)}{4G_N^{(d+1)}}\,,
\label{Sdef}}
where $\gamma_A$ is a co-dimension 2 static minimal surface corresponding to the subsystem $A$ and $G_N^{(d+1)}$ is the $(d+1)$-dimensional Newton's gravitational constant. The quantity computed from \eqref{Sdef} is usually labelled as holographic EE (or HEE, for short) and we will review its computation for several well-known supergravity solutions.

The entanglement entropy is also related to other field theoretical quantities depending on the spacetime dimensions \cite{Ryu:2006ef,Myers:2010tj,Casini:2011kv}. In odd dimensions, the universal piece of entanglement entropy for a spherical subsystem is related to the sphere free energy defined to be (negative of) the logarithm of the partition function of the CFT placed on a $d$-sphere: $S_A=-F_{S^d}≡\log|Z_{S^d}|$. In even dimensions, the universal piece of EE is related to the Weyl anomaly $a_d$ via the relation $S_A=(-1)^{\frac{d}{2}-1}4a_d$. This  captures a part of the trace of the energy-momentum tensor $\langle T_μ^μ\rangle \sim -(-1)^{\frac{d}{2}}2a_d E_d +⋯$, where $E_d$ is the $d$-dimensional Euler density.\footnote{In $d=2$, the $a$-anomaly is related to the central charge $c$ of the 2d CFT so the relation $S_A\sim c$ is more common in this case.} Both the free energy and $a$-anomaly are useful in the study of renormalization group flows \cite{Zamolodchikov:1986gt,Cardy:1988cwa,Komargodski:2011vj,Jafferis:2011zi}. The holographic computations match the corresponding field theoretical results wherever available.
 
\paragraph{Quantum Complexity.} The QC involves minimizing the number of unitary transformations required to transform the state of a system from a reference state to a desired target state. It is a difficult concept to define in a QFT and no satisfactory field theoretical definition of complexity exists yet. But several attempts have been made in field theory to define geometric and circuit complexity \cite{Nielsen:2006qcg,Dowling:2006gqc,Chapman:2017rqy,Jefferson:2017sdb}, and path integral complexity\cite{Caputa:2017yrh,Bhattacharyya:2018wym}.

In addition, there also have been numerous attempts to define the notion of complexity holographically. Initially, it was conjectured that the QC of a state (measured in gates) is proportional to the \emph{volume} of the Einstein-Rosen bridge (ERB) connecting two boundaries of an eternal black hole \cite{Susskind:2014rva,Stanford:2014jda}
\equ{C_{V}(t_L,t_R)=\frac{\mathrm{V_{ERB}(t_L,t_R)}}{8πL_{AdS}G_N^{(d+1)}}\,,
}
where $V_{ERB}$ is defined to be co-dimension 1 maximal volume bounded by the two spatial slices at times $t_L$ and $t_R$ anchored at the entangled state of two CFTs that live on the two boundaries of the eternal black hole. Another proposal for computing QC holographically states that it can be obtained from the bulk \emph{action} evaluated on the Wheeler-DeWitt patch \cite{Brown:2015bva,Brown:2015lvg}
\equ{C_W=\frac{\mathrm{I_{WDW}}}{\pi \hbar}\,·
}
In both these holographic proposals, the complexity depends on the whole state of the physical system at the boundary. The relation between these two ``Complexity=Volume'' and ``Complexity=Action'' conjectures have been explored in detail in \cite{Carmi:2016wjl}, including the generalization to subregion complexity, which involves reducing the boundary state to a specific subregion of the boundary time slice. Following these, there is yet another ``Complexity=Volume'' conjecture proposed in \cite{Alishahiha:2015rta} that specifically depends on the reduced state of the system. This involves computing the maximal co-dimension 1 volume $V(\gamma_A)$ enclosed by the co-dimension 2 static minimal surface $\gamma_A$ (RT surface) foliated into the bulk. Explicitly, it reads
\equ{C_A=\frac{\mathrm{V(\gamma_A)}}{8πL_{AdS}G_N^{(d+1)}}\,,
\label{Cdef}}
where $L_{AdS}$ is the length scale of the AdS space in consideration. This has been dubbed the holographic subregion complexity (HSC) in the literature. We will focus exclusively on this definition in this note to calculate the HSC for a few well-known supergravity solutions containing $AdS_4$--$AdS_7$ spacetimes.

\

We study the relation between HSC and HEE for various supergravity solutions by focussing on their universal pieces\footnote{The universal piece, in this context, refers to a term that does not depend on the chosen subsystem, up to a logarithmic divergence\cite{Ryu:2006ef,Casini:2011kv}.}. We find that for solutions having a product geometry of pure AdS spacetime and a compact manifold, the universal piece of HSC is simply proportional to that of the HEE. However, for solutions having a warped AdS factor (arising due to nontrivial dilaton profile), the application of \eqref{Cdef} does not result in such a simple relation (we show this explicitly in the appendix for two cases), which seems an unlikely result. Because of the fact that both the HEE and HSC are calculated by using the same minimal RT surface, we expect such a simple relation between these two quantities to be a generic feature. Thus, to achieve this, we propose \eqref{Ctdef} as a slight modification of \eqref{Cdef} by arguing that the warp factor needs to be taken into account in defining the AdS length scale $L_{AdS}$ appearing in \eqref{Cdef}.

Most of the solutions we consider have well-known CFT duals and computation of subregion complexity on the gravity side leads to the prediction for the associated quantity on the CFT side, as expected from AdS/CFT correspondence. This also means that the holographic relation between the universal pieces of the HEE and HSC leads to a prediction of a similar relation between the associated CFT observables. Thus, the universal piece of the field theoretical analogue of HSC (which we will denote simply as $C$ in the following sections) can be predicted to be proportional to the $a$-anomaly or the sphere free energy for CFTs in even or odd dimensions, respectively. This fact has not been appreciated in the literature as far as we know, which should lead to a focussed effort in defining and computing the complexity for such dual CFTs, providing further concrete tests of the AdS/CFT correspondence.

\

The rest of this note is organized as follows. In Section \ref{sec:Complexity}, we modify the definition of HSC \eqref{Cdef} to include AdS spacetimes with warp factors. In Sections \ref{sec:10d} and \ref{sec:11d}, we consider a few well-known 10d and 11d supergravity solutions and find a simple relation between universal pieces of the HEE and (modified) HSC. This relation leads to a prediction for field theoretic complexity in terms of either the $a$-anomaly or the free energy on sphere, as discussed above. In the final Section \ref{sec:Discuss} we end with a summary of the results and some future directions. We also include Appendix \ref{app:NaiveHSC} collecting results of straightforward application of the HSC formula \eqref{Cdef}, when it is different from the modified HSC we define below.

\section{Revisiting HSC}\label{sec:Complexity}
The HSC was defined in \cite{Alishahiha:2015rta} by considering pure $AdS_{p+1}$ spacetime. The supergravity solutions, which arise in the weak gravity limit of superstring or M-theory, are product manifolds involving AdS spacetime and a compact manifold. There can also exist nontrivial warp factors for each component of the product manifold. In general, we can consider the following metric (in Einstein frame) for the full $(d+1)$-dimensional spacetime ($d=9$ for string theory and $d=10$ for M-theory):
\equ{ds^2_{d+1}= L_{AdS}^2 F(x)^2ds^2_{AdS_{p+1}} +L_X^2 G(x)^2ds^2_{X_{d-p}}\,,
\label{DefMet}}
where $F(x)$ and $G(x)$ are warp factors multiplying the AdS metric\footnote{We will take the AdS metric to be of the form $ds^2_{AdS_{p+1}}=\frac{1}{z^2}\(-dt^2 +d\vec{x}^2 +dz^2\)$, with $d\vec{x}^2=∑_{i=1}^{p-1}(dx^i)^2=dρ^2+ρ^2ds_{S^{p-2}}^2$.} and metric for the compact manifold $X$, respectively. Such warp factors depending on the $d-p$ coordinates $\{x\}$ of the compact manifold $X$ can arise due to the nontrivial dilaton profile, as we will see later. It is well-known that in the presence of such warp factors, we get an \emph{effective} $AdS$ radius $L_{AdS}(x) = L_{AdS} F(x)$, which can no longer be considered constant when considering the full supergravity solution. This leads to an ambiguity in applying \eqref{Cdef} to evaluate the HSC since it is not clear \emph{which value} of $L_{AdS}$ to use. One way to resolve this ambiguity is to bring $L_{AdS}(x)$ inside the integral defining the $V(γ_A)$ in \eqref{Cdef}. A similar modification was considered in \cite{Macpherson:2014eza} to (re)define central charge defined originally in \cite{Klebanov:2007ws}. Thus, we are led to the following modification of \eqref{Cdef}:
\equ{\widetilde{C}_A=\frac{1}{8πG_N^{(d+1)}}∫_{γ_A}d^{d}x\frac{\sqrt{g^{(d)}}}{L_{AdS}(x)}\,·
\label{Ctdef}}
Here, $g^{(d)}$ denotes the determinant of the $d$-dimensional metric following from \eqref{DefMet} for static surfaces, i.e., $t=0$ in the $AdS_{p+1}$ metric. Also, note that $\wt{C}_A≡C_A$ as given in \eqref{Cdef} when $L_{AdS}(x)$ is constant since $V(γ_A)=∫_{γ_A}d^{d}x\sqrt{g^{(d)}}\,$ and $γ_A$ denotes the RT surface whose area computes the HEE via the relation \eqref{Sdef}.

As discussed in the Introduction, one of the motivation for the modified definition \eqref{Ctdef} is that the universal piece of $\wt{C}_A$ is proportional to that of $S_A$ whereas a direct application of \eqref{Cdef} does not guarantee that (see Appendix \ref{app:NaiveHSC}). Such a simple relation between $S_A$ and $C_A$ is implicit in \cite{Alishahiha:2015rta} and has been further explored in \cite{Momeni:2017ibg,Bhattacharya:2019zkb,Ben-Ami:2016qex} where AdS spacetimes are considered without any explicit embeddings in string or M-theory. We revisit those calculations both for HEE and HSC now in the context of the generic metric with warped AdS factor given in \eqref{DefMet} to prove the proportionality claim.

In order to compute HEE, we consider a subsystem $A$ realized as a round sphere: $\rho^2=∑_{i=1}^{p-1}(x^i)^2\leq R^2$. The embedding of this static ($t=0$) RT surface into the bulk is specified by the profile $\rho=\rho(z)$. The surface area of the RT surface then reads
\equ{\mathrm{Area}(\gamma_A)= ∫d^{d-p}xL_{AdS}(x)^{p-1}L_X(x)^{d-p}\sqrt{g^{(d-p)}}\Vol(S^{p-2}) ∫_{z_0}^{R}dz\frac{ρ(z)^{p-2}\sqrt{1+ρ'(z)^2}}{z^{p-1}}\,·
}
In the above integral we have introduced a UV cut-off $z_0$ to regularize the area functional.\footnote{The $z_0$ can be related to the lattice spacing in the discretized version of the dual field theory\cite{Susskind:1998dq}.} Solving the Euler-Lagrange equation obtained from the above area function we find $\rho(z)= \sqrt{R^2-z^2}$. This leads to the following expression for HEE\cite{Ryu:2006bv,Ryu:2006ef}
\begingroup
\allowdisplaybreaks
\eqs{S_A=\frac{\mathrm{Area}(\gamma_A)}{4G_N^{(d+1)}} & =\frac{\Vol(S^{p-2})}{4G_N^{(d+1)}}∫d^{d-p}xL_{AdS}(x)^{p-1}L_X(x)^{d-p}\sqrt{g^{(d-p)}}∫_{z_0}^{R}dz\,\frac{ρ(z)^{p-2}\sqrt{1+ρ'(z)^2}}{z^{p-1}} \nn
&≈\frac{\Vol(S^{p-2})}{4G_N^{(d+1)}}\I_{(d-p)}×
\begin{cases}
(-1)^{n-1}\Big(\genfrac{}{}{0pt}{1}{n-\frac{3}{2}}{n-1}\Big)\log\big(\frac{2R}{z_0}\big) &\quad p=2n \\[2mm]
\frac{(-1)^n}{2n}\Big(\genfrac{}{}{0pt}{1}{n-\frac{1}{2}}{n}\Big)^{-1} &\quad p=2n+1
\end{cases}\,,
\label{SaGenp}}
\endgroup
where we have denoted the $(d-p)$-dimensional integral as $\I_{(d-p)}$ and kept only the universal piece of the $z$-integral, i.e., $\log$ term for even $p$-dimensional case and constant term for odd one \cite{Ryu:2006ef,Casini:2011kv}.

Now we can compute the HSC for the RT surface specified above using \eqref{Ctdef}
\begingroup
\allowdisplaybreaks
\eqs{\wt{C}_A &=\frac{1}{8πG_N^{(d+1)}}∫_{γ_A}d^{d}x\frac{\sqrt{g^{(d)}}}{L_{AdS}(x)} \nn
&= \frac{1}{8πG_N^{(d+1)}}∫d^{d-p}x\frac{1}{L_{AdS}(x)}L_{AdS}(x)^{p}L_X(x)^{d-p}\sqrt{g^{(d-p)}}\Vol(S^{p-2})∫_{z_0}^{R}dz∫_0^{\sqrt{R^2-z^2}}dρ\,\frac{ρ^{p-2}}{z^p} \nn
&≈\frac{\Vol(S^{p-2})}{8πG_N^{(d+1)}}\I_{(d-p)}×
\begin{cases}
\frac{(-1)^nπ}{2(2n-1)} &\quad p=2n \\[2mm]
\frac{(-1)^n}{2n}\log\big(\frac{R}{z_0}\big) &\quad p=2n+1
\end{cases}\,,
\label{CaGenp}}
\endgroup
where we have again kept only the universal pieces\cite{Alishahiha:2015rta,Ben-Ami:2016qex}. Note that the nature of universal pieces in \eqref{CaGenp} is opposite to those obtained for HEE in \eqref{SaGenp}. It is now straightforward to show that the universal pieces of $S_A$ and $\wt{C}_A$ are proportional independent of the integral over the compact manifold by comparing \eqref{SaGenp} and \eqref{CaGenp}:
\equ{C_p =\begin{cases}
\frac{-1}{4(2n-1)}\Big(\genfrac{}{}{0pt}{1}{n-\frac{3}{2}}{n-1}\Big)^{-1}\frac{\displaystyle S_A}{\log\big(\frac{2R}{z_0}\big)} &\quad p=2n \\[4mm]
\frac{1}{2π}\Big(\genfrac{}{}{0pt}{1}{n-\frac{1}{2}}{n}\Big) S_A &\quad p=2n+1
\end{cases}\,·
\label{CSGenRel}}
Recall that we denote the universal piece of HSC (or, equivalently its field theoretical analogue) simply by $C$ with the subscript $p$ denoting the spacetime dimension of the (dual) CFT. Since these relations are independent of $\I_{(d-p)}$ and $G_N^{(d+1)}$, they do not depend on the explicit embedding in string theory or M-theory and are valid for any generic holographic CFT dual in $p$-dimensions. In this sense, they are ``universal'' relations and we will rewrite them purely from the CFT point of view in Section \ref{sec:Discuss}.

\

We will now show a few explicit examples of the above relations in the following sections for some well-known supergravity solutions.

\section{String Theory Solutions}\label{sec:10d}
In this section, we study the relation between HEE and HSC of the 10-dimensional supergravity solutions of the form $AdS_5×X_5$ and $AdS_6×Y_4$ and what that entails for associated field theoretical quantities.

\subsection[\texorpdfstring{$AdS_5\times  X_5$}{AdS₅×X₅}]{$\bm{AdS_5\times  X_5}$}\label{sec:AdS5X5}
The AdS$_5$/CFT$_4$ is the most well-studied AdS/CFT correspondence. Many 4d $\N≥1$ SCFTs have been constructed that have type IIB string theory duals on $AdS_5×X_5$, where $X_5$ is a compact 5-dimensional Sasaki-Einstein manifold \cite{Maldacena:1997re,Witten:1998qj,Klebanov:1998hh}. The 10d supergravity metric in general reads
\equ{ds^2= L^2\left[\frac{-dt^2+d\vec{x}^2 +dz^2}{z^2}\right] +L^2ds^2_{X_5}\,,
\label{eqMetAdS5}}
where $d\vec{x}^2=∑_{i=1}^3(dx^i)^2=dρ^2+ρ^2ds_{S^2}^2$ with $\Vol(S^2)=4π$. The self-dual 5-form flux quantization relation is given by
\equ{\frac{L^4}{l_s^4}=\frac{4π^4N}{\Vol(X_5)}\,·
\label{LlsRel}}
We will also need
\equ{G_N^{(10)}=\frac{(2πl_s)^8}{32π^2}
}
relating the 10d gravitational constant to string length $l_s$.

We follow the generic calculation done in the previous section to compute HEE here. That is, we consider a spherical subsystem $A$ given by $\rho^2=∑_{i=1}^3(x^i)^2\leq R^2$, whose embedding into the bulk is $\rho=\rho(z)= \sqrt{R^2-z^2}$. This leads to the following expression for HEE:
\begingroup
\allowdisplaybreaks
\eqs{S_A=\frac{\mathrm{Area}(\gamma_A)}{4G_N^{(10)}} & =\frac{8π^2L^8}{(2πl_s)^8}\Vol(X_5)\Vol(S^2)∫_{z_0}^{R}dz\,\frac{ρ(z)^2\sqrt{1+ρ'(z)^2}}{z^3} \nn
&≈\frac{2π^3N^2}{\Vol(X_5)^2}\Vol(X_5)\left[-\frac{1}{4} -\frac{1}{2}\log\(\frac{2R}{z_0}\) +\frac{R^2}{2z_0^2} +\O(z_0^2)\right] \nn
&≈-\frac{π^3N^2}{\Vol(X_5)}\log\(\frac{2R}{z_0}\)·
\label{Sa5d}}
\endgroup
The coefficient of the $\log$ term is the universal piece, which is equal to the 4d Weyl anomaly as follows
\equ{\frac{S_A}{\log\big(\frac{2R}{z_0}\big)}=-4a_{\text{4d}} \qquad ⇒\qquad a_{\text{4d}}=\frac{π^3N^2}{4\Vol(X_5)}\,·
}
Note the $N^2$ dependence and that matches the $a$-anomaly at large $N$ for 4d SCFTs \cite{Gubser:1998vd,Gubser:1998fp}.

Now, we proceed to compute the volume enclosed by the embedding RT surface, which is given by
\equ{V(\gamma_A) =L^{9}\Vol(X_5)\Vol(S^2)∫_{z_0}^{R}dz∫_0^{\sqrt{R^2-z^2}}dρ\frac{ρ^2}{z^4}\,·
}
Since $L_{AdS}=L$ is a constant, $\wt{C}_A=C_A$ and the HSC can be easily evaluated to be
\begingroup
\allowdisplaybreaks
\eqs{\wt{C}_A =\frac{V(\gamma_A)}{8πLG_N^{(10)}} & =\frac{L^9(4π)}{L(2πl_s)^8}\Vol(X_5)(4π)∫_{z_0}^{R}dz∫_0^{\sqrt{R^2-z^2}}dρ\frac{ρ^2}{z^4} \nn
&≈\Vol(X_5)\frac{π^2N^2}{\Vol(X_5)^2}\left[\frac{π}{6} +\frac{R^3}{9z_0^3} -\frac{R}{2z_0} +\O(z_0)\right] \nn
&≈\frac{π^3N^2}{6\Vol(X_5)}\,·
}
\endgroup
We again keep only the universal piece in the last step, which is the $R$-independent term here. Comparing it with \eqref{Sa5d}, we obtain a relation between the 4d Weyl anomaly and $C_{\text{4d}}$:
\equ{\wt{C}_A=-\frac{1}{6}\frac{S_A}{\log\big(\frac{2R}{z_0}\big)} \qquad ⇒ \qquad C_{\text{4d}}=\frac{2}{3}a_{\text{4d}}\,.
}

\subsection[\texorpdfstring{$AdS_6\times Y_4$}{AdS₆×Y₄}]{$\bm{AdS_6\times Y_4}$}\label{sec:AdS6Y4}
The 5d $\N=1$ SCFTs have seen a lot of activity recently and have been engineered in both type IIA and IIB string theory. One of the simplest class of 5d SCFTs is that of Seiberg theories whose gravity duals are given by massive type IIA string theory on $AdS_6×S^4$\cite{Seiberg:1996bd,Brandhuber:1999np,Bergman:2012kr}. The 10d supergravity metric in string frame explicitly reads
\equ{ds^2= \frac{L^2}{(\sin α)^{\frac{1}{3}}}\left[\frac{-dt^2+d\vec{x}^2 +dz^2}{z^2}\right] +\frac{4L^2}{9(\sin α)^{\frac{1}{3}}}\(dα^2+\cos^2α\,ds^2_{S^3/\bZ_n}\),
\label{5dMet}}
where $d\vec{x}^2=∑_{i=1}^4(dx^i)^2=dρ^2+ρ^2ds_{S^3}^2$ with $\Vol(S^3)=2π^2$ and $α∈(0,\frac{π}{2}]$. The dilaton and 4-form flux quantization relation are given by
\eqs{e^{-2φ} &=\frac{3(8-N_f)^{\frac{3}{2}}\sqrt{nN}}{2\sqrt{2}π}(\sin α)^{\frac{5}{3}} \label{eqDilaton}\\
\frac{L^4}{l_s^4} &=\frac{18π^2nN}{8-N_f}\,·
}

We again choose a spherical subsystem $A$ and following the RT prescription, we find the entanglement entropy\footnote{This calculation is to be done in Einstein frame, so we need to use $g_{μν}^{E}→e^{-\frac{φ}{2}}g_{μν}^s ⇒ \sqrt{g^{(8),E}} → e^{-2φ}\sqrt{g^{(8),s}}$.} \cite{Jafferis:2012iv}
\begingroup
\allowdisplaybreaks
\eqs{S_A &=\frac{\mathrm{Area}(\gamma_A)}{4G_N^{(10)}}=\frac{2}{(2π)^6l_s^8}∫d^8x\,e^{-2φ}\sqrt{g^{(8)}} \nn
&=\frac{(8-N_f)^{\frac{3}{2}}\sqrt{N}}{3^3\sqrt{2n}π^3}\left[\frac{18π^2nN}{8-N_f}\right]^2∫_0^{\frac{π}{2}}dα\sin^{\frac{1}{3}}α\cos^3α∫_{z_0}^{R}dz\frac{ρ(z)^3\sqrt{1+ρ'(z)^2}}{z^4} \nn
&=\frac{3×4π n^{\frac{3}{2}}N^{\frac{5}{2}}}{\sqrt{2(8-N_f)}}\frac{9}{20}\left[\frac{2}{3} -\frac{R}{z_0}+\frac{R^3}{3z_0^3}\right] \nn
&≈\frac{9\sqrt{2}π n^{\frac{3}{2}}N^{\frac{5}{2}}}{5\sqrt{8-N_f}}\,·
\label{SA5d}}
\endgroup
We again keep the universal piece in the last step. The above result satisfies the relation $S_A=-F_{S^5}$, where $F_{S^5}$ is the $S^5$ free energy of the Seiberg theories, as shown in \cite{Jafferis:2012iv}.

We now compute HSC using the modified definition \eqref{Ctdef} here\footnote{See Appendix \ref{app:NaiveHSC} for naive application of the definition \eqref{Cdef}.} because, as is clear from the metric \eqref{5dMet} and dilaton profile \eqref{eqDilaton}, the AdS radius is not constant but depends on the $α$ coordinate of the compact manifold as follows (in Einstein frame):
\equ{L_{AdS}(x)=\frac{L}{\sin^{\frac{1}{6}}α}e^{-\frac{φ}{4}}\,.
}
This leads to the same large $N$ scaling for HSC as that of HEE:
\begingroup
\allowdisplaybreaks
\eqs{\widetilde{C}_A &=\frac{1}{8πG_N^{(10)}}∫d^9x\frac{e^{-\frac{9}{4}φ}\sqrt{g^{(9)}}}{L(\sin^{-\frac{1}{6}}α)e^{-\frac{φ}{4}}} =\frac{2}{(2π)^7L l_s^8}∫d^9x e^{-2φ}\sin^{\frac{1}{6}}α\sqrt{g^{(9)}} \nn
&=\frac{(8-N_f)^{\frac{3}{2}}\sqrt{N}}{3^3\sqrt{2n}2π^4}\left[\frac{18π^2nN}{8-N_f}\right]^2∫_0^{\frac{π}{2}}dα\sin^{\frac{1}{3}}α\cos^3α∫_{z_0}^Rdz∫_0^{\sqrt{R^2-z^2}}dρ\frac{ρ^3}{z^5} \nn
&=\frac{6n^{\frac{3}{2}}N^{\frac{5}{2}}}{\sqrt{2(8-N_f)}}\frac{9}{20}\left[\frac{3}{16} +\frac{1}{4}\log\(\frac{R}{z_0}\) -\frac{R^2}{4z_0^2}+\frac{R^4}{16z_0^4}\right] \nn
&≈\frac{27\sqrt{2}n^{\frac{3}{2}}N^{\frac{5}{2}}}{80\sqrt{8-N_f}}\log\(\frac{R}{z_0}\)·
}
\endgroup
We have again kept the universal piece in the last step, which gives the expected relation between free energy and $C_{5d}$:
\equ{\frac{\widetilde{C}_A}{\log\big(\frac{R}{z_0}\big)} =\frac{3}{16π}S_A \qquad ⇒ \qquad C_{\text{5d}}=-\frac{3}{16π}F_{S^5}\,.
}

\section{M-theory Solutions}\label{sec:11d}
In this section, we again verify that the HEE and HSC are proportional for SCFTs with well-known supergravity duals arising in the weak gravity limit of M-theory and discuss what that means for the corresponding field theoretical quantities.

\subsection[\texorpdfstring{$AdS_4×Y_7$}{AdS₄×Y₇}]{$\bm{AdS_4×Y_7}$}\label{sec:AdS4Y7}
The AdS$_4$/CFT$_3$ correspondence was put on a concrete footing after the discovery of $\N=6$ ABJM theory \cite{Aharony:2008ug} describing the low energy limit of a stack of $N$ M2-branes placed at the tip of cone over $S^7/\bZ_k$. In the large $N$ limit, ABJM theory is dual to M-theory on $AdS_4×S^7/\bZ_k$. After this discovery, a large number of 3d $\N≥2$ SCFTs with M-theory duals have been identified by replacing $S^7/\bZ_k$ with $Y_7$, a compact (tri-)Sasaki-Einstein 7-manifold. Following \cite{Marino:2011nm}, we can write the general metric for the 11d supergravity solution as
\equ{ds^2= \frac{L^2}{4}\left[\frac{-dt^2+d\vec{x}^2 +dz^2}{z^2}\right] +L^2ds^2_{Y_7}\,,
}
where $d\vec{x}^2=∑_{i=1}^2(dx^i)^2=dρ^2+ρ^2dθ^2$ with $0≤θ<2π$ and the 4-form flux quantization condition that relates the geometric length scale $L$ to Planck length $l_p$:
\equ{\frac{L^6}{l_p^6}=\frac{(2π)^6N}{6\Vol(Y_7)}\,·
}
We will also use the relation of 11d gravitational constant to $l_p$:
\equ{G_N^{(11)} = \frac{(2\pi l_p)^9}{32\pi^2}\,·
\label{Gtol}}

Following the RT prescription for a spherical subsystem $A$, we find for HEE
\begingroup
\allowdisplaybreaks
\eqs{S_A=\frac{\mathrm{Area}(\gamma_A)}{4G_N^{(11)}} & =\frac{2}{(2π)^7}\frac{L^9}{l_p^9}∫_0^{2π}dθ∫_{z_0}^{R}dz\Vol(Y_7)\frac{ρ(z)\sqrt{1+ρ'(z)^2}}{(2z)^2} \nn
&≈\frac{\Vol(Y_7)}{2(2π)^6}\left[\frac{(2π)^6N}{6\Vol(Y_7)}\right]^{\frac{3}{2}}\left[-1+\frac{R}{z_0}\right] \nn
&≈-\frac{\sqrt{2}π^3N^{\frac{3}{2}}}{3\sqrt{3\Vol(Y_7)}}\,,
\label{sM1}}
\endgroup
where we keep only the universal piece ($R$-independent term) in the last step.
It is a well-known fact that the HEE as given in \eqref{sM1} matches the $S^3$ free energy of the dual SCFTs in the large $N$ limit via $S_A=-F_{S^3}$\cite{Klebanov:1996un,Drukker:2010nc,Herzog:2010hf}.

Now, we proceed to compute the volume enclosed by the embedding RT surface, which is given by
\equ{V(\gamma_A) =2\pi L^{10} ∫_{z_0}^{R}dz∫_0^{\sqrt{R^2-z^2}}dρ\Vol(Y_7)\frac{ρ}{(2z)^3}\,·
}
Since $L_{AdS}=\frac{L}{2}$, we have $\wt{C}_A=C_A$ and so the HSC turns out to be
\eqs{\wt{C}_A =\frac{V(\gamma_A)}{8π\(\frac{L}{2}\)G_N^{(11)}} 
&≈\frac{\Vol(Y_7)}{2(2π)^7}\left[\frac{(2π)^6N}{6\Vol(Y_7)}\right]^{\frac{3}{2}}\left[-\frac{1}{4} -\frac{1}{2}\log\(\frac{R}{z_0}\) +\frac{R^2}{4z_0^2}\right]\nn
&≈-\frac{\sqrt{2}π^2N^{\frac{3}{2}}}{12\sqrt{3\Vol(Y_7)}}\log\(\frac{R}{z_0}\)·
}
Note that $\wt{C}_A$ also scales as $N^{\frac{3}{2}}$ just like $S_A$. In this case, the universal piece is the coefficient of the logarithmic term and hence, comparing it with $S_A$, we get the following relation:
\equ{\frac{\wt{C}_A}{\log\big(\frac{R}{z_0}\big)}=\frac{1}{4π}S_A\qquad ⇒\qquad C_{\text{3d}}=-\frac{1}{4π}F_{S^3}\,.
}
The above relation implies that in the large $N$ limit, $C_{3d}$ is proportional to the $S^3$ free energy for the 3d SCFTs having M-theory duals.

\subsection[\texorpdfstring{Uplift of NATD of $AdS_5\times S_5$}{Uplift of NATD of AdS₅×S₅}]{Uplift of NATD of $\bm{AdS_5\times S_5}$}\label{sec:natdAdS5S5}
Let us now consider the M-theory uplift of the solution obtained by applying nonabelian T-duality (NATD) to $AdS_5×S^5$.\footnote{We consider here only the case of $S^5$. Other cases discussed in \cite{Macpherson:2014eza} yield similar results as one can verify. More generic backgrounds have also been considered in \cite{Nunez:2019gbg} along with their CFT duals. It would be interesting to compute their HSC explicitly.} The details are in \cite{Sfetsos:2010uq,Macpherson:2014eza} and we collect here only the relevant expressions including the 11d metric
\eqst{ds^2= e^{-\frac{2}{3}Φ}ds^2_{AdS_5} +e^{\frac{4}{3}Φ}\(dy-2\frac{L^4\cos^4α}{α'^{\frac{3}{2}}}dθ\)^2 \\
+e^{-\frac{2}{3}Φ}\left[4L^2\(dα^2+\sin^2αdθ^2\)+\frac{{α'}^2dβ^2}{L^2\cos^2α} +\frac{e^{2Φ}L^4β^2\cos^4α\(dξ^2\sin^2χ +dχ^2\)}{α'}\right],
}
where we use the $AdS_5$ metric given in \eqref{eqMetAdS5} and $e^{-2Φ}=\frac{L^2}{{α'}^3}\cos^2α\(L^4\cos^4α +{α'}^2β^2\)$. The flux quantization condition gives the following relation
\equ{L^4=2^{\frac{8}{γ}}N^{\frac{2}{γ}}{α'}^2\,,
}
where $γ$ is introduced by scaling the coordinate $y→(\frac{L^2}{α'})^γ\sqrt{α'}\,y$ due to an ambiguity in the uplifting procedure. We will also use (only in this subsection) $G_N^{(11)}={α'}^{\frac{9}{2}}$ following \cite{Macpherson:2014eza}, relating the 11d gravitational constant to string tension $α'$.

To compute HEE, we again consider a spherical subsystem and following the RT prescription, we have the surface area integral given by
\eqst{\mathrm{Area}(\gamma_A)= 4L^8 \Vol(S^2)∫_{z_0}^{R}dz∫_0^{2π}dy\,dθ\,dξ∫_0^πdβ\,dχ∫_0^{\frac{π}{2}}dα \\
×\(\frac{L^2}{α'}\)^γ\sqrt{α'}β^2\cos^3α\sin α\sin χ\frac{ρ(z)^2 \sqrt{1+ρ'(z)^2}}{z^3}\,·
}
Again, setting $\rho(z)= \sqrt{R^2-z^2}$ as in the previous examples, we get for HEE
\begingroup
\allowdisplaybreaks
\eqs{S_A=\frac{\mathrm{Area}(\gamma_A)}{4G_N^{(11)}} & =\frac{L^8\sqrt{α'}}{{α'}^{\frac{9}{2}}}\(\frac{L^2}{α'}\)^γ\frac{4π^6}{3}\Vol(S^2)∫_{z_0}^{R}dz\,\frac{ρ(z)^2\sqrt{1+ρ'(z)^2}}{z^3} \nn
&≈\frac{2^{8(1+\frac{2}{γ})}π^7N^{1+\frac{4}{γ}}}{3}\left[-\frac{1}{4} -\frac{1}{2}\log\(\frac{2R}{z_0}\) +\frac{R^2}{2z_0^2} +\O(z_0^2)\right] \nn
&≈-\frac{2^{8(1+\frac{2}{γ})}π^7N^{1+\frac{4}{γ}}}{6}\log\(\frac{2R}{z_0}\)·
\label{SAnatd}}
\endgroup
We have kept the universal piece in the last step, which should equal the 4d Weyl anomaly:
\equ{\frac{S_A}{\log\big(\frac{2R}{z_0}\big)}=-4a_{\text{4d}} \qquad ⇒\qquad a_{\text{4d}}=\frac{2^{8(1+\frac{2}{γ})}π^7N^{1+\frac{4}{γ}}}{24}\,·
}
Note that $a_{\text{4d}}=\frac{π}{8}c$, where $c$ is the central charge for this solution obtained in \cite{Macpherson:2014eza}. For $γ=4$, we have the usual $N^2$ scaling of 4d and for $γ=2$, we have $N^3$ scaling reminiscent of 6d, that we will see in the next example.

Now, we compute the HSC but since the AdS radius $L_{AdS}(x)=e^{-\frac{1}{3}Φ}$ is coordinate dependent, we use the modified definition of complexity \eqref{Ctdef} to obtain\footnote{See Appendix \ref{app:NaiveHSC} for the naive result from the definition \eqref{Cdef}.}
\begingroup
\allowdisplaybreaks
\eqs{\wt{C}_A &=\frac{1}{8πG_N^{(11)}}4L^8\sqrt{α'}\(\frac{L^2}{α'}\)^γ\frac{16π^7}{3}∫_{z_0}^{R}dz∫_0^{\sqrt{R^2-z^2}}dρ\frac{ρ^2}{z^4} \nn
&≈\frac{2^{8(1+\frac{2}{γ})}π^6N^{1+\frac{4}{γ}}}{6}\left[\frac{π}{6} +\frac{R^3}{9z_0^3} -\frac{R}{2z_0} +\O(z_0)\right] \nn
&≈\frac{2^{8(1+\frac{2}{γ})}π^7N^{1+\frac{4}{γ}}}{36}\,·
}
\endgroup
We again keep only the universal piece ($R$-independent term) in the last step. Comparing it with \eqref{SAnatd}, we obtain a relation between the 4d Weyl anomaly and $C_{4d}$:
\equ{C_A=-\frac{1}{6}\frac{S_A}{\log\big(\frac{2R}{z_0}\big)} \qquad ⇒ \qquad C_{\text{4d}}=\frac{2}{3}a_{\text{4d}}\,.
}
This is the same relation that we got for the $AdS_5×X_5$ solution. In fact, this relation is ``universal'' for $AdS_5$ and is independent of the uplift to either string theory or M-theory as expected from the general discussion of Section \ref{sec:Complexity}.

\subsection[\texorpdfstring{$AdS_7\times X_4$}{AdS₇×X₄}]{$\bm{AdS_7\times X_4}$}\label{sec:AdS7X4}
The 6d SCFTs are strongly interacting non-Lagrangian theories describing the low energy limit of $N$ M5-branes. At large $N$, the $\N=(2,0)$ SCFTs are dual to M-theory on $AdS_7×S^4/Γ$, where the compact manifold $X_4$ can only be an orbifold of the 4-sphere $S^4$ with $Γ$ being a discrete subgroup of $SU(2)$\cite{Maldacena:1997re,Ferrara:1998vf}.\footnote{The $\N=(1,0)$ SCFTs also have interesting M-theory duals with warped $AdS_7$ factors (see \cite{Filippas:2019puw} and references therein). They satisfy the same ``universal'' relation but we do not consider such metrics here.} The metric of this 11d supergravity solution explicitly reads
\equ{ds^2= L^2\left[\frac{-dt^2+d\vec{x}^2 +dz^2}{z^2}\right] +\frac{L^2}{4}ds^2_{S^4/Γ}\,,
}
where $d\vec{x}^2=∑_{i=1}^5(dx^i)^2=dρ^2+ρ^2ds_{S^4}^2$ with $\Vol(S^4)=\frac{8π^2}{3}\,·$ The 4-form flux quantization relation is given by
\equ{\frac{L^3}{l_p^3}=8π|Γ|N\,·
}

Similar to the previous examples, we choose a spherical geometry of the subsystem $A$ with the profile of the corresponding RT surface being $ρ(z)=\sqrt{R^2-z^2}$, which leads to
\begingroup
\allowdisplaybreaks
\eqs{S_A=\frac{\mathrm{Area}(\gamma_A)}{4G_N^{(11)}} & =\frac{2}{(2π)^7}\frac{L^9}{l_p^9}∫_{z_0}^{R}dz\Vol(S^4)\frac{\Vol(S^4/Γ)}{2^4}\frac{ρ(z)^4\sqrt{1+ρ'(z)^2}}{z^5} \nn
&≈\frac{1}{8(2π)^7}\frac{8π^2}{3}\frac{8π^2}{3|Γ|}\big[8π|Γ|N\big]^3\left[\frac{9}{32} +\frac{3}{8}\log\(\frac{2R}{z_0}\) -\frac{3R^2}{4z_0^2} +\frac{R^4}{4z_0^4} +\O(z_0^2)\right] \nn
&≈\frac{4N^3|Γ|^2}{3}\log\(\frac{2R}{z_0}\),
\label{sm2}}
\endgroup
where we keep only the universal piece in the last step with the famous $N^3$ scaling \cite{Klebanov:1996un}. The coefficient of the $\log$ term in $S_A$ is proportional to the 6d Weyl anomaly:
\equ{\frac{S_A}{\log\big(\frac{2R}{z_0}\big)}=4a_{\text{6d}} \qquad ⇒\qquad a_{\text{6d}}=\frac{1}{3}N^3|Γ|^2\,.
}
This matches the $a$-anomaly at large $N$ for 6d SCFTs, at least the $N^3|Γ|^2$ factor \cite{Henningson:1998gx,Bastianelli:2000hi,Cordova:2015fha}.\footnote{The exact coefficient seems to depend on a ``scheme-dependent'' definition of the 6d Euler density, or equivalently, the choice of renormalization of the anomaly contribution of the free $\N=(2,0)$ tensor multiplet. We do not attempt to fix this coefficient here.}

Now, we can compute the complexity ($\wt{C}_A=C_A$ here) following steps similar to the previous examples and it reads
\begingroup
\allowdisplaybreaks
\eqs{C_A =\frac{V(\gamma_A)}{8πLG_N^{(11)}} & =\frac{2}{(2π)^8}\frac{L^{10}}{L l_p^9}\frac{8π^2}{3}\frac{8π^2}{3|Γ|}\frac{1}{2^4}∫_{z_0}^{R}dz∫_0^{\sqrt{R^2-z^2}}dρ\frac{ρ^4}{z^6} \nn
&≈\frac{1}{9·2(2π)^4|Γ|}\big[8π|Γ|N\big]^3\left[-\frac{π}{10} +\frac{R^5}{25z_0^5} -\frac{R^3}{6z_0^3} +\frac{3R}{8z_0} +\O(z_0)\right] \nn
&≈-\frac{8N^3|Γ|^2}{45}\,·
}
\endgroup
We have again kept only the universal piece in the last step and comparing with the $S_A$ result in \eqref{sm2}, we obtain
\equ{C_A=-\frac{2}{15}\frac{S_A}{\log\big(\frac{2R}{z_0}\big)} \qquad ⇒\qquad C_{\text{6d}}=-\frac{8}{15}a_{\text{6d}}\,.
}
The above relation implies that $C_{6d}$ is proportional to the 6d $a$-anomaly in the large $N$ limit.

\section{Discussion}\label{sec:Discuss}
We have obtained holographic subregion complexity by computing the co-dimension 1 maximal volume enclosed by the co-dimension 2 Ryu-Takayanagi surface in $AdS_{p+1}$ with $p=3,4,5,6$ for specific supergravity solutions, most of which are known to have explicit SCFT duals. We found that the universal piece of HSC is proportional to that of HEE calculated holographically via the RT prescription for those AdS backgrounds without warp factors, as has been expected in the literature.\footnote{The time-dependent analysis of complexity in $AdS_3$ \cite{Ageev:2018nye,Ageev:2019fxn} shows that the linear relation between HEE and HSC holds only at initial times with the relation evolving into quite a nontrivial one at later times. It would be interesting to generalize such an analysis to higher dimensions and extract some ``universal'' behaviour.} However, we observe that in case of gravity duals with nontrivial warp factors (due to a nontrivial dilaton profile) modifying the AdS part of the supergravity backgrounds, the expected proportionality between the universal pieces of HSC and HEE does not hold anymore. In order to retain this simple relation, we propose a modification of the holographic formula to compute complexity as explained in Section \ref{sec:Complexity}. The existence of a warp factor implies that there is an effective non-constant $L_{AdS}(x)$ depending on the warp factor, leading us to the modified definition of complexity in \eqref{Ctdef}. This simple fact drastically affects the computation of the volume enclosed by the co-dimension 2 RT surface, as one can contrast the calculations of HSC using \eqref{Cdef} in the Appendix \ref{app:NaiveHSC} with those using \eqref{Ctdef} in Sections \ref{sec:10d} and \ref{sec:11d}.

The relation between the universal pieces of HEE and HSC is of great importance as it enables us to predict the behavior of the corresponding field theoretical quantity. We find that at large $N$, in odd dimensional CFTs, the universal piece of the field theoretical analogue of HSC ($C_p$) is proportional to the sphere free energy $F_{S^p}$, whereas for even dimensional CFTs, it is proportional to the Weyl $a$-anomaly. We can write a general relation for these quantities, as it straightforwardly follows from \eqref{CSGenRel} and the relation of $S_A$ to $F_{S^p}$ or $a$-anomaly \cite{Casini:2011kv}:
\equ{C_p =\begin{cases}
\frac{-1}{2π}\Big(\genfrac{}{}{0pt}{1}{n-\frac{1}{2}}{n}\Big) F_{S^p} &\quad p=2n+1 \\[3mm]
\frac{(-1)^n}{2n-1}\big(\genfrac{}{}{0pt}{2}{n-\frac{3}{2}}{n-1}\big)^{-1} a_p &\quad p=2n
\end{cases}\,.
}
Note that these relations hold irrespective of the explicit nature of the dual gravity theory whether embedded in string theory or M-theory. We take this ``universal'' relation (for a given $p$, of course) as a justification for the modification we propose for the holographic prescription to compute HSC.

\

Even though, a satisfactory and universal definition of complexity in field theory is lacking at present, the definition involving path integral optimization\cite{Caputa:2017yrh,Bhattacharyya:2018wym} seems to be promising as it could lead to application of localization techniques for computing complexity. These techniques have been remarkably successful in obtaining exact results for $F$'s and $a$'s in SCFTs in various dimensions\cite{Pestun:2016zxk}, which we used to compare holographic results in the large $N$ limit. Another set of ``universal'' relations can be obtained between field theoretic complexities across dimensions by employing the results of \cite{Bobev:2017uzs}. For example, $C_{\text{3d}}=-\frac{32}{27}(\fg-1)C_{\text{5d}}$, given that $F_{S^3}=-\frac{8}{9}(\fg-1)F_{S^5}$ for 5d theories defined on $S^3×Σ_{\fg}$ with a topological twist on $Σ_{\fg}$\cite{Crichigno:2018adf}.

\

We also note that our analysis was restricted to subsystems defined by spherical surfaces. But there have also been considerable interest in singular surfaces as these lead to the appearance of more universal pieces including $\log²$ behaviour, as discovered in \cite{Myers:2012vs,Bakhshaei:2017qud,Bakhshaei:2019ope,Bueno:2019mex}. It would be interesting to revisit the relation between such universal pieces of HEE and HSC explicitly in this context.

\

It is also worth mentioning that many proposals have been given which relate the HSC with other information theoretical quantities, like the Fisher information metric and the Bures metric (fidelity susceptibility)\cite{Miyaji:2015mia,Alishahiha:2017cuk,Banerjee:2017qti,Karar:2019bwy}. These are standard notions of distances in quantum information theory\cite{Braunstein:1994zz,Wootters:1981ki,Safranek:2017onq} and arise holographically when one considers an excitation of the dual spacetime geometry. For example, the Fisher information metric is defined to be the difference between the RT volumes of the excited geometry and background geometry, considering up to second order perturbation about the background geometry. In this paper, we have considered only pure $AdS$ geometries without any excitations so it would be interesting to see how the second order variations of HSC and HEE relate to each other and whether one can still relate these metrics to well-studied calculable properties like the free energy or $a$-anomaly of the dual CFTs.

\section*{Acknowledgements}
SG acknowledges the support of the Visiting Associateship programme of Inter University Centre for Astronomy and Astrophysics (IUCAA), Pune. AS would like to acknowledge the support by Council of Scientific and Industrial Research (CSIR, Govt. of India) for a Senior Research Fellowship. The authors also acknowledge the anonymous referee for very useful comments.

\newpage
\appendix
\section{Naive Complexity Calculations}\label{app:NaiveHSC}
This appendix collects the computation of HSC using the expression \eqref{Cdef} for the examples in Subsections \ref{sec:AdS6Y4} and \ref{sec:natdAdS5S5} with nontrivial warp factors leading to different large $N$ scaling compared to HEE. This, in part, led us to modify \eqref{Cdef} to the expression given in \eqref{Ctdef}.

\subsection[\texorpdfstring{$AdS_6\times Y_4$}{AdS₆×Y₄}]{$\bm{AdS_6\times Y_4}$}
Here is the result one would get by naively using the formula \eqref{Cdef} to compute the HSC:
\begingroup
\allowdisplaybreaks
\eqs{C_A &=\frac{V(\gamma_A)}{8πLG_N^{(10)}} =\frac{2}{(2π)^7L l_s^8}∫d^9x e^{-\frac{9}{4}φ}\sqrt{g^{(9)}}\nn
&=\frac{(8-N_f)^{\frac{27}{16}}N^{\frac{9}{16}}}{18×2^{\frac{11}{16}}×3^{\frac{7}{8}}π^{\frac{33}{8}}n^{\frac{7}{16}}}\left[\frac{18π^2nN}{8-N_f}\right]^2∫_0^{\frac{π}{2}}dα\sin^{\frac{3}{8}}α\cos^3α∫_{z_0}^R dz∫_0^{\sqrt{R^2-z^2}}dρ\frac{ρ^3}{z^5} \nn
&=\frac{2^{\frac{5}{16}}3^{\frac{9}{8}}n^{\frac{25}{16}}N^{\frac{41}{16}}}{π^{\frac{1}{8}}(8-N_f)^{\frac{5}{16}}}\frac{128}{297}\left[\frac{3}{16} +\frac{1}{4}\log\(\frac{R}{z_0}\) -\frac{R^2}{4z_0^2}+\frac{R^4}{16z_0^4}\right] \nn
&≈\frac{32×2^{\frac{5}{16}}n^{\frac{25}{16}}N^{\frac{41}{16}}}{33×3^{\frac{7}{8}}π^{\frac{1}{8}}(8-N_f)^{\frac{5}{16}}}\log\(\frac{R}{z_0}\)·
}
\endgroup
We kept the universal piece in the last line, which has a different large $N$ scaling as compared to $S_A$ in \eqref{SA5d}.

\subsection[\texorpdfstring{Uplift of NATD of $AdS_5\times S_5$}{Uplift of NATD of AdS₅×S₅}]{Uplift of NATD of $\bm{AdS_5\times S_5}$}
The following is the result obtained by naively using the formula \eqref{Cdef} to compute the HSC:
\begingroup
\allowdisplaybreaks
\eqs{C_A &=\frac{V(\gamma_A)}{8πLG_N^{(11)}} =\frac{L^8\sqrt{α'}}{2π{α'}^{\frac{9}{2}}}\(\frac{L^2}{α'}\)^γ(4π)∫_0^{2π}dy\,dθ\,dξ∫_0^πdβ\,dχ \nn
&\hspace*{3cm}×∫_0^{\frac{π}{2}}dα\,e^{-\frac{1}{3}Φ}β^2\cos^3α\sin α\sin χ ∫_{z_0}^{R}dz∫_0^{\sqrt{R^2-z^2}}dρ\frac{ρ^2}{z^4} \nn
&≈2×2^{8(1+\frac{2}{γ})}π^3N^{1+\frac{4}{γ}}(2^{\frac{4}{γ}}N^{\frac{1}{γ}})^{\frac{1}{2}}∫_0^{\frac{π}{2}}dα∫_0^πdββ^2\cos^{4}α\sin α\left[\frac{π}{6} +\frac{R^3}{9z_0^3} -\frac{R}{2z_0} +\O(z_0)\right] \nn
&≈\frac{2^{2(4+\frac{9}{γ})}π^7N^{1+\frac{9}{2γ}}}{45}\,·
}
\endgroup
We again have the universal piece in the last line with a different large $N$ scaling when compared to $S_A$ in \eqref{SAnatd}.

\vfill
\references{bibcomp}

\end{document}